# Deterministic One-to-One Synthesis of Germanium Nanowires and Individual Gold Nano-Seed Patterning for Aligned Nanowire Arrays**

Dunwei Wang, Ryan Tu, Li Zhang and Hongjie Dai[*]

* D. Wang, R. Tu, L. Zhang and Prof. Dr. H. Dai

Department of Chemistry

Stanford University

Stanford, CA 94305 (USA)

Fax: (+1) 650-725-0259.

Tel: (+1) 650-723-4518.

Email: Hongjie Dai (hdai@stanford.edu)

** This work was supported by a DARPA 3D Electronics Program, the Stanford INMP and SRC/AMD. R. T. acknowledges a NSF graduate fellowship.

Germanium nanowires (GeNWs) have attracted much attention in recent years[1-9] owing to the advanced electrical properties of Ge such as high carrier mobilities[10] and the ability of facile chemical synthesis of single crystal GeNWs at low temperatures well below 400 °C[5]. Various synthetic methods for crystalline GeNWs have been reported using Au nanoparticles as seeds, including chemical vapor deposition (CVD) [5] [8] [9] [11], physical vapor deposition (PVD) [2, 3] and solvo-thermal reactions[1] [4]. Nevertheless, one-to-one growth (1-1 growth), i.e. each Au particle seed producing one nanowire[12] and the chemistry needed for such growth has not been well-established thus far for GeNWs. Achieving this goal is important to the deterministic synthesis of nanowires with controlled yield, size, growth locations and orientations.

Here, we demonstrate gaining excellent control over the synthesis of GeNWs by low temperature CVD. First, we achieve 100% yield of GeNWs relative to the Au nanoparticle seeds. By understanding and optimizing the growth chemistry, we achieve 1-1 correspondence of GeNWs with Au seeds. We uncover an interesting phenomenon that the optimum growth temperature for GeNWs is dependent on the size of the Au nanoparticle. Secondly, we develop a method capable of patterning individual Au nanoparticles and grow GeNWs with 100% yield from the well-defined nanoparticle arrays. Lastly, we show that the orientation of the GeNWs grown from the patterned sites can be manipulated by a post-growth flow aligning treatment to afford quasi-parallel arrays of GeNWs with well defined spacing.

GeNW synthesis in the current work is based on a previous CVD method[5] with a modification in that a low pressure CVD (LPCVD) system (Fig. 1a) is used to replace the previous atmosphere pressure CVD (APCVD) one. GeNWs are grown at various temperatures in the range of 270-400°C on substrates decorated with preformed Au nanoparticles (diameter in the range of 5-50 nm) using $GeH_4$ as the precursor for Ge feedstock. As shown previously, the growth mechanism is well described by the vapor-liquid-solid (VLS) model[5, 13, 14]. We find that a main advantage of LPCVD synthesis over APCVD is that $GeH_4$ concentration in the CVD system is better controlled by varying the pressure in the system than diluting $GeH_4$ with carrier gases. We identify the partial pressure of $GeH_4$ for optimum GeNW growth is between 4-8 Torr, below which the yield of NWs is low due to insufficient feedstock and above which undesirable pyrolysis of $GeH_4$ is observed. Another important advantage of the LPCVD approach is the rapid removal of $O_2$ and $H_2O$ species trapped in the system by vacuum pumping. This efficiently reduces the contaminants in the system and makes the growth results highly reproducible between experiments.

The main growth result and understanding obtained by the current work is that the optimum GeNW growth condition is size dependent. That is, the growth temperatures at which optimum 1-1 growth of GeNWs can be achieved vary with the sizes of Au seeds. Under a fixed partial pressure of $GeH_4$ of $P$=5 Torr, the optimum growth temperature for d=20±2 nm Au seeds is around 295°C under which every Au seed can produce a GeNW (Fig. 1b and c). It can be seen that the total number of nanowires grown matches with the number of starting Au nanoparticle seeds, and the nanowires are originated from the positions of the starting particles (Fig. 1b). The 100% yield and 1-1 growth at 295°C for the 20 nm particles are robust and have been reproduced with 10 batches of samples. For larger d=50±3 nm Au seeds, the optimum GeNW growth temperature is ~ 310 °C at which 1-1 growth can be achieved (Fig. 2b). At a lower temperature of 295°C, not all of the 50±3 nm Au seeds are capable of producing GeNWs (Fig. 2a). On the other hand, if the growth temperature is high (at e.g., 325°C), the 50nm Au seeds are found to produce more GeNWs than the number of starting seed particles and interestingly, NWs with diameters much smaller than the starting d~50 nm particles are observed (Fig. 2c). This observation suggests that the ~ 50 nm Au seeds have split into smaller ones to produce smaller NWs at the relatively high growth temperature (Fig. 2d).

The optimum GeNW CVD growth temperatures for various size Au seed particles in the range of 5nm to 50nm are summarized in Fig. 3. A general trend is that smaller Au seeds can nucleate and grow nanowires at lower temperatures. For large particles (d~50 nm), low growth temperature under-produce GeNWs with low yield and too high a temperature tends to over-produce wires due to splitting of Au seeds. These results can be explained by considering several key factors involved in the VLS growth process. The first factor is that the eutectic melting temperature of Ge-Au is size dependent and higher for larger particles. Such size dependence of melting temperature has been documented for single and binary element particles[15] [16]. It is therefore reasonable that larger particles require higher temperature for efficient supersaturation and growth to occur. Secondly, Ge diffusion in the Au particle is an important kinetic factor of the VLS growth process. The size of the Au seeds determines the diffusion length over which Ge must reach to saturate the Ge-Au solution for nucleation and growth of one nanowire from the seed particle. Higher temperatures will facilitate Ge diffusion and thus NW growth from larger particles. The third factor is Ge feedstock supply. Higher temperature will lead to more efficient decomposition of $GeH_4$ precursor and provide an efficient Ge supply need for larger Au particles.

The VLS growth of NWs from large Au seed particles appears to be diffusion limited. At high temperatures, the feeding of feedstock could be rapid while the diffusion of feedstock atoms in Au might not be sufficiently high to supersaturate a large particle. Rather, smaller regions of the Au cluster are supersaturated rapidly, leading to nucleation and growth of smaller NWs from the parent Au particle (Fig. 2d). As control experiments, we have attempted growth of Ge and Si NWs (GeNWs using $GeH_4$ and SiNWs using $SiH_4$) from ultra large Au particles with d~250nm. Under all experimental temperatures tested, we are unable to achieve 1-1 growth from these large particles and always observed small wire growth due to particle splitting. We believe that the diffusion limitation for large NW growth, and the size dependent NW growth is general to the synthesis of various NW materials via the VLS mechanism.

With the 1-1 growth ability, we next pursue patterning of individual Au nanoparticles to achieve 1-1 growth of GeNWs at controlled locations with monodispersed sizes. This is achieved by first using electron-beam lithography to pattern arrays of small Au islands on a substrate (Fig. 4a). The islands are 40 nm on the side with various thickness in the range of 1-10 nm formed by the evaporation and liftoff technique. Upon annealing at 300 ºC, Au atoms in each island are found to aggregate and form well-defined Au dots with controllable diameters in the range of 5-50 nm (dot size dependent on the metal thickness in the 40 nm wide islands). Fig. 4b shows an array of regularly spaced d~20±3 nm Au particles formed by this method. CVD growth using the optimum condition identified in Fig.3 for 20 nm Au seeds leads to successful 1-1 growth of GeNWs from the nanoparticle arrays (Fig.5). This result demonstrates that NW synthesis can be well controlled at the single particle level by making use of the understanding of nanowire growth and state-of-the-art lithographic patterning technique.

We have also explored controlling the orientations of the GeNWs.  Our approach is to utilize fluid flow[17] [18] to manipulate and re-orient the GeNWs grown from the patterned Au particle arrays, as shown schematically in Fig. 5a.  Due to the VLS tip-growth process, one of the ends of an as-grown GeNW is anchored on the substrate (at where the nanowire is grown from as highlighted by arrows in Fig. 5b and d) and can act as a pivotal point for the wire.  After a stream of DI $H_2O$ is flowed across the substrate surface, we find that the nanowires are re-oriented towards the flow direction and become quasi-aligned while maintaining the same spacing between their pivoted ends (Fig. 5c and e).

To summarize, we have demonstrated controlled one-to-one synthesis of GeNWs with ~100% yield from Au seed particles.  Size dependent optimum GeNW growth conditions are identified.  For $\leq 50$ nm Au seeds, under-growth, 1-1 growth and over-growth can occur depending on the growth conditions. Growth of large GeNWs appears to be diffusion limited. These results should have generic implications to the synthesis of other types of nanowires via VLS.  Patterning and positioning of individual Au nanoparticles are achieved by lithographic patterning and used for successful one-to-one nanowire growth.  Finally, post-growth flow-alignment is used to obtain quasi-parallel nanowires originating from well-controlled locations. These results are important to the fundamental science of nanomaterials synthesis and may find important applications in various fields including high performance nanoelectronics.

**Experimental Section**

**Deposition of pre-formed Au nanoparticles**. A silicon substrate is soaked in 3-aminopropyltriethylsilane (APTES) aqueous solution (12μL APTES in 20mL $H_2O$) for 2 minutes. After thorough rinsing with DI $H_2O$ and blow-drying with $N_2$, the substrate is then soaked in Au colloid solution (sizes of 5-50nm) for 5 minutes. The resulting density of Au particles on the substrate is determined by colloidal solution concentration. For instance, 20nm Au colloids solutions purchased from Ted Pella, Inc. (CA, USA) contain $10^{11}$ nanoparticles/mL. 1000-time dilution is performed to obtain a concentration of $10^8/cm^3$. When deposited on a silicon substrate, a density of ~1 particles/$3\mu m^2$ can be reliably obtained with such a concentration. After Au nanoparticle deposition, calcination in air at 300°C for 15 minutes is carried out to remove organic residues. The as-prepared substrate is imaged under scanning electron microscope (SEM, FEI XL30 Sirion) to record the locations of individual nanoparticles in certain regions. Then, the substrate is subject to CVD growth followed by SEM imaging of the same regions to correlate the synthesized nanowires with their parental nanoparticle seeds.

**CVD synthesis of GeNWs.** In a typical growth, the CVD quartz tube-chamber (1-inch) is first evacuated to its base pressure of 150mTorr and then heated up to a growth temperature in the range of 270-325°C. Afterwards, the chamber is filled with precursor species of $GeH_4$ (Germane, 10% in He, Voltaix Inc. NJ, USA) to desired growth pressure (~50Torr total pressure, $GeH_4$ partial pressure ~5Torr) and kept at that pressure throughout the growth. During this process, $GeH_4$ is flowed at a rate of 10sccm (standard cubic centimeter per minute). At the end of the reaction, the feeding of $GeH_4$ is stopped and the chamber is pumped to its base pressure again, followed by cooling down to room temperature. One of the criteria of optimum growth is that after CVD, visual inspections should find that the quartz growth chamber is free of pyrolytic deposits of $GeH_4$ during growth.

**Patterning of individual Au nanoparticles**. A 100 nm thick polymethylmethacrylate (PMMA) film is formed by spin-coating on a Si substrate. Electron beam lithography (Raith 150) is used to create wells in the PMMA film with dimensions of 40×40nm. A 1-10nm thick Au film is then deposited on the PMMA patterned substrate in an e-beam evaporator followed by lift-off of the PMMA to afford 40nm wide Au islands. The substrate is then annealed in Ar at 300°C for 15 minutes during which the small Au islands aggregate to form single Au nanoparticles.

**Manipulating the orientations of nanowires after one-to-one growth.** After CVD growth of GeNWs on a substrate with patterned Au dots, a $H_2O$ droplet is placed onto the substrate to cover the as-grown GeNWs. A $N_2$ flow is then passed to blow-dry the surface along a desired direction. After this simple process, we find that the nanowires can be quasi-aligned with the flow direction.

**Figure Captions:**

**Figure 1.** One-to-one growth of 20 nm GeNWs. a) Schematic of the LPCVD setup. b) SEM of six d~20 nm Au nanoparticles (labeled 1 to 6) as growth seeds. c) Six GeNWs grown from the Au seeds in b). Inset: HRTEM image of a single-crystalline GeNW grown under the same condition as that in the main panel.

**Figure 2.** Growth results from 50 nm Au seeds at a) 295°C, b) 310°C and c) 325°C respectively. Several 'blobs' in a) correspond to Au particles failed to produce nanowires. 1-1 growth is achieved in b) suggesting optimum nanowire growth at 310°C for 50 nm seeds. The lower panel in c) is a atomic force microscopy height profile of GeNWs grown at 325°C showing that in addition to 50nm GeNWs, there are also smaller (d~20 nm) nanowires grown. d) A schematic illustration of multiple nanowires grown from one large parent particle.

**Figure 3.** Size dependent nanowire growth conditions. Optimum nanowire growth temperature vs. the size of Au particle seeds by CVD under a fixed partial pressure of germane of 5 Torr. For each particle size, the optimum growth temperature is denoted by a hollow square and the error bars spans less than 10 degrees.

**Figure 4.** Patterning of individual Au nanoparticles. a) A schematic of the patterning process. b) An atomic force microcopy (AFM) image of an array of Au dots regularly

spaced at 2 μm. c) A zoom-in AFM image of one Au nanoparticle. d) Height profile of the Au nanoparticle in c) showing the size of the particle ~20nm.

**Figure 5.** Post-growth re-orientation of nanowires. a) A schematic illustration of the process to flow-align GeNWs grown from an array of patterned Au seeds. b) A SEM image of GeNWs grown from an array of 20 nm Au dots similar to that in Fig. 4b. The arrows indicate the original positions of the Au dots and the fixed ends of the as-grown GeNWs. c) SEM of quasi-aligned GeNWs (2μm pitch) after flow treatment of the sample in b). d) and e) SEM images of GeNWs 1-1 grown from patterned Au dots (5 μm pitch) before and after flow-align treatment.

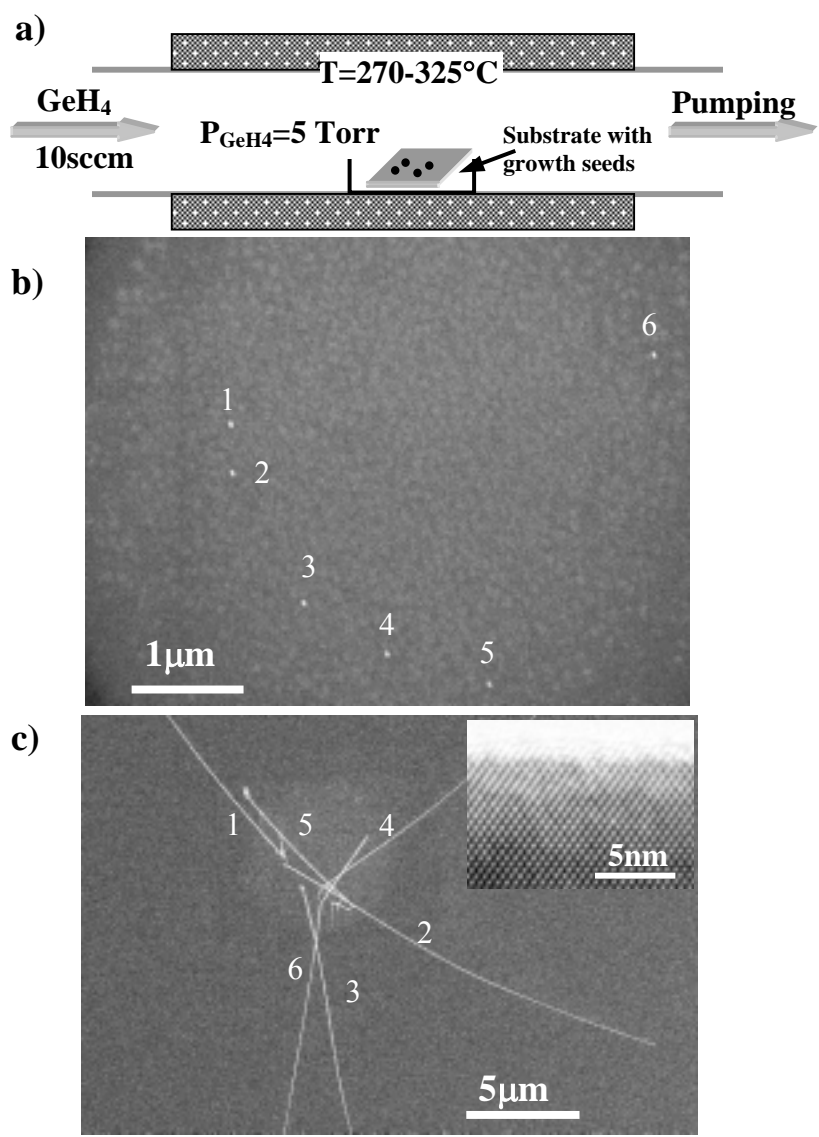

**a)** T=270-325°C

GeH₄
10sccm   $P_{GeH4}$=5 Torr   Substrate with growth seeds   Pumping

**b)** 6

4   5

1µm

**c)** 1   5   4

5nm

6   3

5µm

**Figure 1**

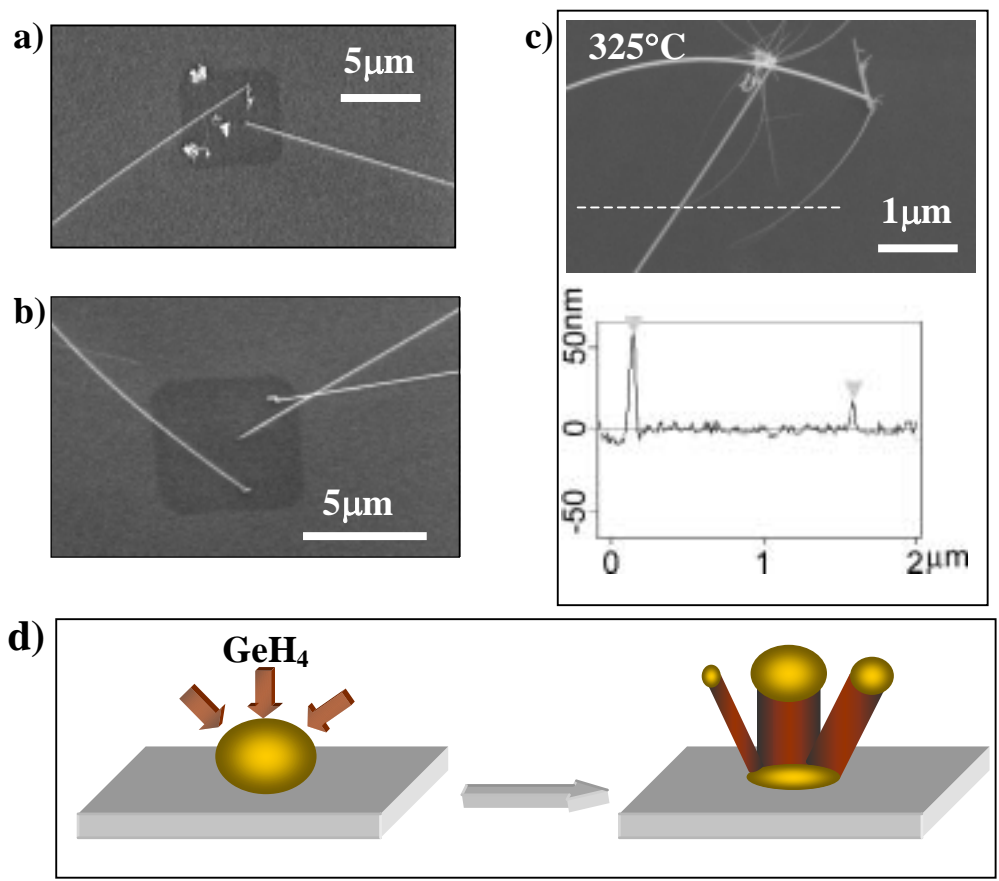

**Figure 2**

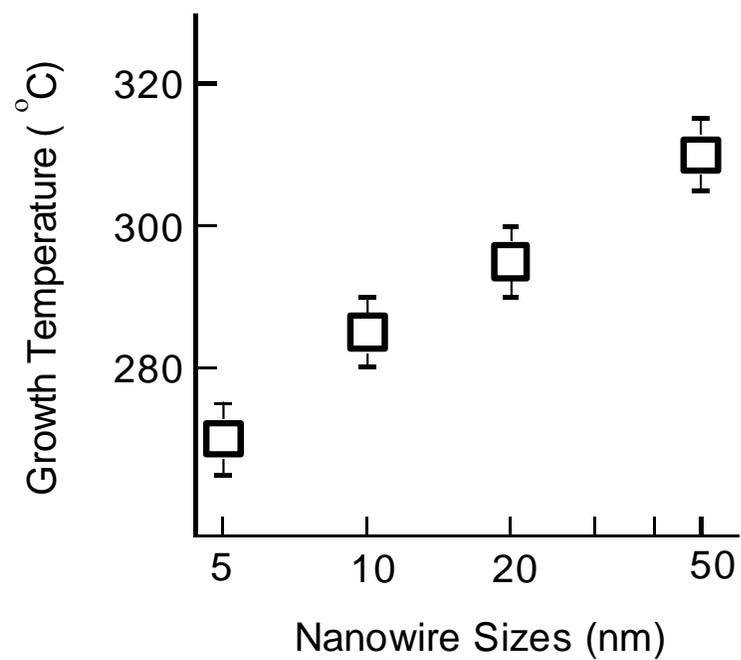

**Figure 3**

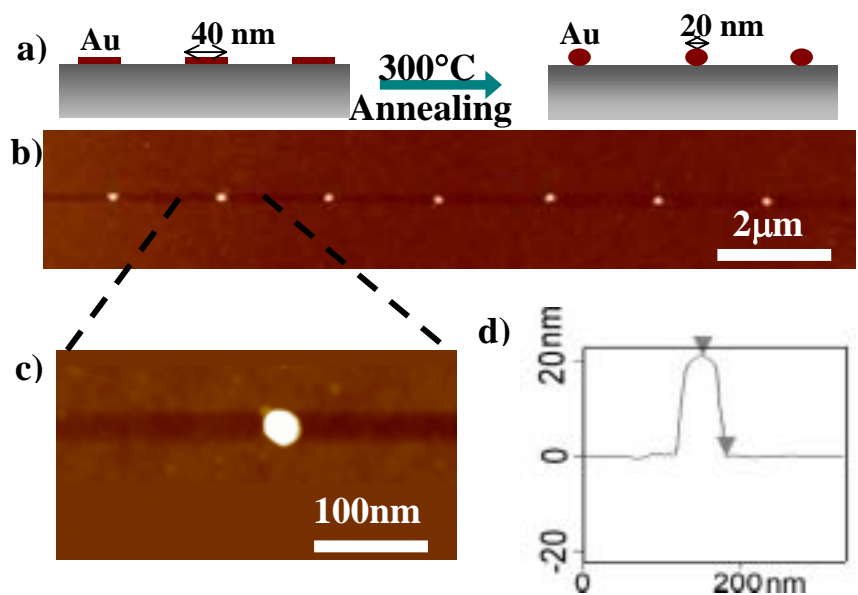

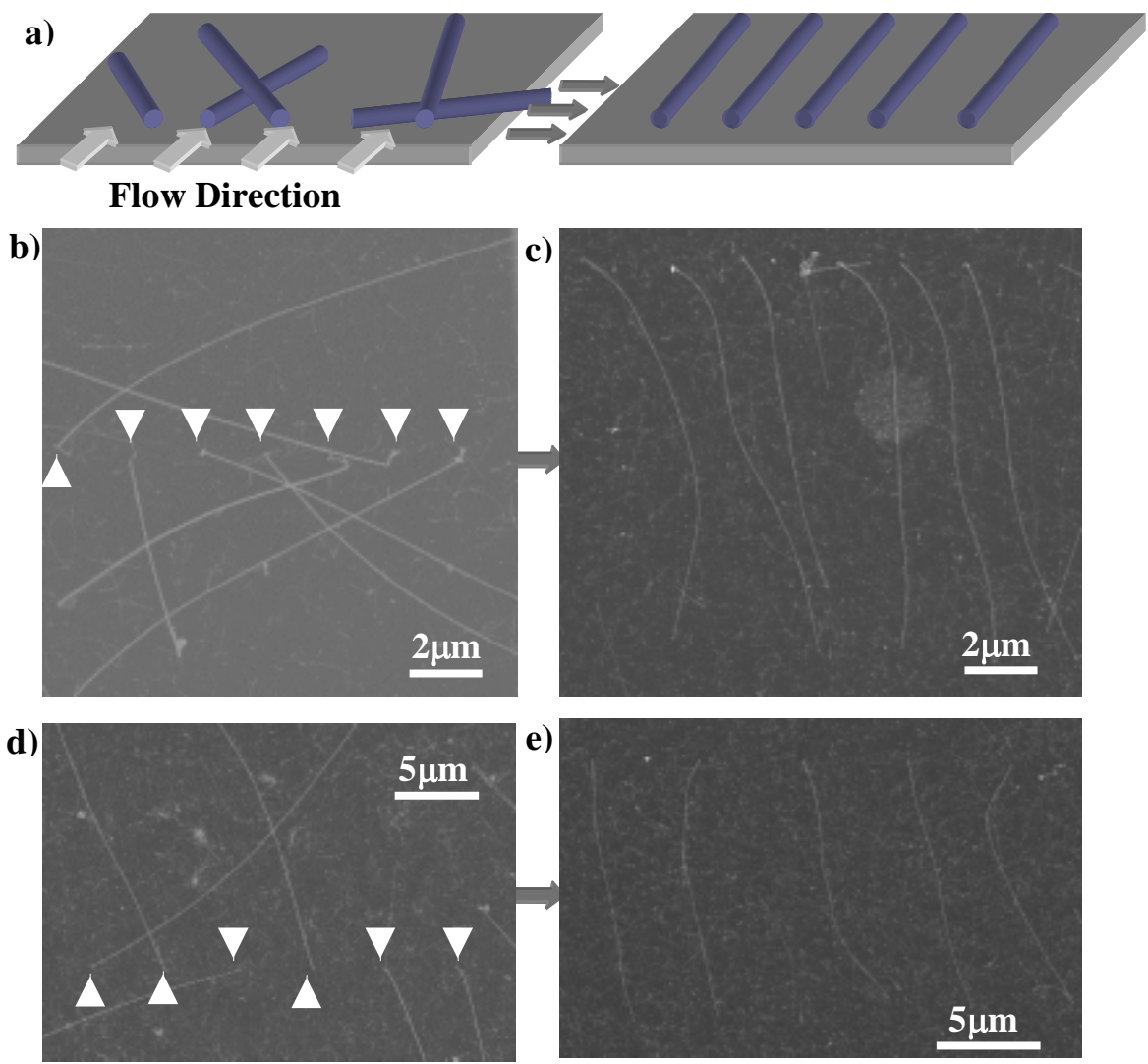

**Figure 5**